\documentclass{ws-ijgmmp}

\begin{document}

\markboth{SVS, ACK \& AHH}
{A Study of  Morris-Thorne Wormhole in Einstein-Cartan Theory }

\catchline{}{}{}{}{}
%
%%%%%%%%%%%%%%%%%%%%%%%%%%%%%%%%%%%%%%%%%%%%%%%%%%%%%%%%%%%%%%%%%%%%

\title{A Study of  Morris-Thorne Wormhole in Einstein-Cartan Theory 
}
\author{Sagar V. Soni \footnote{Corresponding author.}}

\address{Department of Mathematics, Sardar Patel University\\
Vallabh Vidyanagar, Gujarat 388120, India.\\
\email{sagar.soni7878@gmail.com
} }

\author{A. C. Khunt}

\address{Department of Physics, Sardar Patel University\\
Vallabh Vidyanagar, Gujarat 388120, India.\\
\email{ankitkhunt@spuvvn.edu
} }

\author{A.H. Hasmani}

\address{Department of Mathematics, Sardar Patel University\\
Vallabh Vidyanagar, Gujarat 388120, India.\\
\email{ah\_hasmani@spuvvn.edu
} }

\maketitle

\begin{history}
\received{(Day Month Year)}
\revised{(Day Month Year)}
\end{history}

\begin{abstract}
This paper focuses on the Einstein-Cartan theory, an extension of general relativity that incorporates a torsion tensor into spacetime. The differential form technique is employed to analyze the Einstein-Cartan theory, which replaces tensors with tetrads. A tetrad formalism, specifically the Newmann-Penrose-Jogia-Griffiths formalism, is used to study the field equations. The energy-momentum tensor is also determined, considering a Weyssenhoff fluid with anisotropic matter. The spin density is derived in terms of the red-shift function. We also examine the energy conditions at the throat of a Morris-Thorne wormhole. The results shed light on the properties of wormholes in the context of the Einstein-Cartan theory, including the energy conditions at the throat.
\end{abstract}

\keywords{Wormhole, Differential Forms, Newmann-Penrose-Jogia-Griffiths Formalism, Energy conditions, Einstein-Cartan Theory }

\section{Introduction}	
In 1915, Einstein devised a theory called general relativity, in which he demonstrated that gravitation is a geometric property rather than a force. Later, in 1924, Cartan \cite{cartan_24} extended this theory by incorporating a torsion tensor into spacetime, which is known as Einstein-Cartan theory. Because of the torsion tensor, spacetime does not possess in Riemannian geometry, hence people in this field work in non-Riemannian geometry. This non-Riemannian aspect of spacetime is described by affine connection $\tilde{\Gamma}^h_{ij}$ and it is defined as,
\begin{align}\label{1.1}
\tilde{\Gamma}^h_{ij}&=\Gamma^h_{ij}-K_{ij}\hspace{0.1mm}^h,
\end{align}
where $\Gamma^h_{ij}$ are the usual Christoffel symbols, which hold symmetry property in lower indices and the symbols $K_{ij}\hspace{0.1mm}^h$ are the contortion tensor with $K_{i(jh)}=0$, that is skew-symmetric in last two indices. The torsion tensor in terms of contortion tensor is given by
\begin{align}\label{1.2}
Q_{ij}\hspace{0.1mm}^h&=-\frac{1}{2}(K_{ij}\hspace{0.1mm}^h-K_{ji}\hspace{0.1mm}^h).
\end{align}
It is worth noting that the torsion tensor is nothing but the anti-symmetric part of affine connection (\ref{1.1}) and hence from the (\ref{1.2}) the following relations hold
\begin{align}\label{1.3}
K_{ij}\hspace{0.1mm}^h=Q_j\hspace{0.1mm}^h \hspace{0.1mm}_i-Q^h\hspace{0.1mm}_{ij}-Q_{ij}\hspace{0.1mm}^h.
\end{align}
\indent In all the modified theories of relativity, researchers are interested to find exact solutions to field equations and analyze energy conditions in different spacetimes. Kibble \cite{kib_61} discovered spin and torsion in gravitation separately. Hehl. et. al. \cite{hehl_73, hehl_74, hehl_76} created the Einstein-Cartan theory of gravitation. In 1939, Tolman \cite{tol_39} developed an explicit solution to Einstein's field equations for static fluid spheres. Prasanna \cite{pr_75} presented the solution's with the perfect fluid distribution in 1975, using Hehl's approach and Tolman's technique, and discovered that a space-time metric similar to the Schwarzschild interior solution will no longer represent a homogeneous fluid sphere in the presence of spin density, and the hydrostatic pressure is discontinuous at the fluid sphere's boundary. Jogia and Griffiths \cite{jog_80} extended the Newmann-Penrose formalism technique for Einstein-Cartan theory in 1980. In 2009, Katkar \cite{katkar_9} used differential forms to derive Cartan's equations of structure, Einstein field equations, and Bianchi's identities. The exact solution of Einstein-Cartan field equations for static, conformally flat spherically symmetric space-time have been derived by Katkar and Patil \cite{katkar_12}. Katkar \cite{katkar_15} used differential forms to determine the general observer values in 2015. In the framework of the Einstein-Cartan theory, Bronnikov and Galiakhmetov \cite{bro_15} investigated the possibility of static traversable wormholes without the use of exotic matter. Di Grazia et. al. \cite{di_17} calculated the torsion tensor for matter fields with varying spins for traversable wormholes. Mehdizadeh and Ziaie \cite{me_17, me_19} have discovered wormhole solutions. Katkar and Phadatare \cite{katkar_191, katkar_192} have established a solution for non-static conformally flat spherically symmetric spacetimes and static spherically symmetric spacetimes with the source as a Weyssenhoff fluid in 2019. Recently a consistent solution of Einstein–Cartan equations with torsion outside matter has been found by Morawetz \cite{mor_21}. Raja et. al. \cite{sahoo1_2023} have extensively elaborated a particular choice of shape functions in the regime of $f(R, L_m)$ gravity. They have explored physical analysis using the isotropic fluid equation of state (EoS) as well as anisotropic fluid EoS, to investigate the physical plausibility of wormhole solutions in the framework of $f(R, L_m)$ gravity.\\
\indent In this paper, we have obtained expressions of Ricci tensor, Ricci scalar, and energy-momentum tensor in tetrad frame using  Newmann-Penrose-Jogia-Griffiths formalism and Katkar's approach in differential forms. Spin density has been derived in the form of a red-shift function. Also, we analyze energy conditions at the throat of the wormhole.Based on the Raja et. al. \cite{sahoo1_2023} wormhole study, we adopted one particular choice of shape function. In this study, we shall explore geometrical as well physical conditions for wormholes in the framework of Einstien-Cartan theory. 

\section{Einstein-Cartan theory in Differential Forms}
The differential form technique \cite{isrl} is immensely used in Einstein's general relativity for numerous calculations, particularly in finding exact solutions. In this tool, tetrads are used instead of tensors. There are many approaches for dealing with tetrads, such as Newmann-Penrose formalism, Geroch-Held-Penrose formalism, and so on, among these, Newmann-Penrose formalism \cite{np_62} is the most commonly used formalism. McIntosh \cite{mh85} derives the relationship between Newmann-Penrose formalism and differential forms.  This formalism was extended by Jogia and Griffiths \cite{jog_80} to Einstein-Cartan theory. This formalism was later known as the Newmann-Penrose-Jogia-Griffiths formalism.  The basis 1-forms and usual basis  are related by,
\begin{align}\label{1}
\theta^\alpha&= e^{(\alpha)}\hspace{0.1mm}_idx^i,
\end{align}
where,  $e^{(\alpha)}\hspace{0.1mm}_i$, represent basis vectors of the Newmann-Penrose tetrad consists of real and complex null vector fields provided by
\begin{align}\label{2}
e^{(\alpha)}\hspace{0.1mm}_i&=(n_i, l_i, -\bar{m}_i, -m_i).
\end{align}
The Greek letters denote tetrad indices, whereas the Latin indices denote tensor indices; all of these indices go from 1 to 4, and this nomenclature will be used throughout the study. Einstein's summation convention is also employed. The metric tensor field is expressed as,
\begin{align}
    g_{ij}=\eta_{(\alpha) (\beta)} e^{(\alpha)}\hspace{0.1mm}_i e^{(\beta)}\hspace{0.1mm}_j
\end{align}
where $\eta_{(\alpha) (\beta)}=e^i\hspace{0.1mm}_{(\alpha)} e_{(\beta)j}$.
The Newmann-Penrose complex null vector fields $l_i, n_i, m_i, \bar{m}_i$ are satisfying conditions,
\begin{align}\label{3}
l_in^i=1=-m_i\bar{m}^i,
\end{align}
and
\begin{align}\label{4}
l_il^i=n_in^i=m_im^i=\bar{m_i}\bar{m}^i=0,\nonumber\\
m_il^i=\bar{m}_il^i=m_in^i=\bar{m_i}n^i=0.
\end{align}
Cartan's first equation of structure is provided by Katkar \cite{katkar_9, katkar_15} in the Einstein-Cartan theory are given by
\begin{align}\label{4}
d\theta^\alpha=-\omega^\alpha\hspace{0.1mm}_\beta \wedge \theta^\beta,
\end{align}
where
\begin{align}\label{5}
\omega^\alpha\hspace{0.1mm}_\beta=(\gamma^\alpha\hspace{0.1mm}_{\beta\delta}-K_{\delta\beta}\hspace{0.09cm}^{\alpha})\theta^\delta
\end{align}
are connection 1-forms which depend on torsion also. Here $\gamma^\alpha\hspace{0.1mm}_{\beta\delta}$ are Ricci roatation coefficients and $K_{\delta\beta}\hspace{0.09cm}^{\alpha}$ are tetrad components of contorsion tensor. The covariant form equations (\ref{5}) can be written as,
\begin{align}\label{6'}
    \omega_{\alpha \beta}= \eta_{\alpha \epsilon} \omega^\epsilon\hspace{0.1mm}_\beta.
\end{align}
The non-vanishing tetrad components of connection 1-forms are represented using notations devised by Jogia and Griffiths and from equation (\ref{6'}) as
\begin{align}\label{6}
\omega_{12}&=-[(\epsilon+\bar{\epsilon}+\epsilon_1+\bar{\epsilon}_1)\theta^1+(\gamma+\bar{\gamma}+\gamma_1+\bar{\gamma}_1)\theta^2 \nonumber \\&\hspace{0.5cm}+(\bar{\alpha}+\beta+\bar{\alpha_1}+\beta_1)\theta^3+(\alpha+\bar{\beta}+\alpha_1+\bar{\beta}_1)\theta^4],\nonumber\\
\omega_{13}&=-[(\kappa+\kappa_1)\theta^1+(\tau+\tau_1)\theta^2+(\sigma+\sigma_1)\theta^3+(\rho+\rho_1)\theta^4],\nonumber\\
\omega_{23}&=(\pi+\pi_1)\theta^1+(\bar{\nu}+\bar{\nu}_1)\theta^2+(\bar{\lambda}+\bar{\lambda_1})\theta^3+(\bar{\mu}+\bar{\mu}_1)\theta^4 ,\nonumber\\
\omega_{34}&=(\epsilon-\bar{\epsilon}+\epsilon_1-\bar{\epsilon}_1)\theta^1+(\gamma-\bar{\gamma}+\gamma_1-\bar{\gamma}_1)\theta^2 \nonumber \\&\hspace{0.5cm}-(\bar{\alpha}-\beta+\bar{\alpha_1}-\beta_1)\theta^3+(\alpha-\bar{\beta}+\alpha_1-\bar{\beta}_1)\theta^4,
\end{align}
where the symbols $\epsilon, \kappa, \nu, ...$ indicate the Ricci rotation coefficients in Einstein's theory of relativity and the symbols with subscript demonstrates contortion tensor.  \\
In the Einstein-Cartan theory, Cartan's second equation of structure is,
\begin{align}\label{7}
\Omega^\alpha\hspace{0.1mm}_\beta&=d\omega^\alpha\hspace{0.1mm}_\beta+\omega^\alpha\hspace{0.1mm}_\sigma\wedge\omega^\sigma\hspace{0.1mm}_\beta+(\gamma^\alpha\hspace{0.1mm}_{\beta\sigma}-K_{\sigma\beta}\hspace{0.01cm}^\alpha)K_{\epsilon\delta}\hspace{0.01cm}^\alpha\theta^\delta\wedge\theta^\epsilon
\end{align}
where
\begin{align}\label{8}
\Omega^\alpha\hspace{0.1mm}_\beta=-\frac{1}{2}R_{\delta\epsilon\beta}\hspace{0.011cm}^\alpha\theta^\delta\wedge\theta^\epsilon
\end{align}
are the components of curvature 2-forms, they give tetrad components of the Riemann-Christoffel tensor. 
\subsection{Field Equations in Einstein-Cartan Theory}
Hehl et. al. \cite{hehl_74, hehl_76} provide the field equations for the Einstein-Cartan theory of gravitation in the following format:
\begin{align}\label{9}
R_{ij}-\frac{1}{2}Rg_{ij}=-Kt_{ij},
\end{align}
and
\begin{align}\label{10}
Q_{ij}\hspace{0.01cm}^k+\delta_i^k Q_{jl}\hspace{0.01cm}^l-\delta_k^j Q_{il}\hspace{0.01cm}^l=kS_{ij}\hspace{0.01cm}^k,
\end{align}
here $t_{ij}$ is energy-momentum tenosr and $S_{ij}\hspace{0.01cm}^k$ is the spin angular momentum tensor. The spin angular momentum tensor $S_{ij}\hspace{0.01cm}^k$ was decomposed by Hehl et al. \cite{hehl_74, hehl_76} into the spin density tensor $S_{ij}$ by,
\begin{align}\label{11}
S_{ij}\hspace{0.01cm}^k=S_{ij}u^k,
\end{align}
with Frankel's condition, That is,
\begin{align}\label{12}
S_{ij}u^j=0.
\end{align}
The spin density tensor $S_{ij}$ comprises six distinct components and is anti-symmetric in nature. These components can be defined in tetrad components as,
\begin{align}\label{13}
S_0&=S_{13}=S_{ij}l^im^j,\nonumber\\
S_1&=\frac{1}{2}(S_{12}-S_{34})=\frac{1}{2}S_{ij}(l^in^j-m^i\bar{m}^j), \nonumber\\
S_2&=S_{32}=S_{ij}m^in^j
\end{align}
and the other three are complex conjugates of the above three components. Therefore, $S_{ij}$ can be written in the form of tetrad as,
\begin{align}\label{14}
S_{ij}=2[(S_1+\bar{S}_1)l_{[i}n_{j]}+(S_1-\bar{S}_1)m_{[i}\bar{m}_{j]}+(\bar{S}_2l_{[i}m_{j]}+\bar{S}_0m_{[i}n_{j]})+C.C.],
\end{align}
where $C.C,$ denotes the complex conjugate of preceding terms.
The condition (\ref{12}) gives,
\begin{align}\label{14'}
S_0=S_2, \hspace{0.5cm} S_1=-\bar{S_1}.
\end{align}
The equation (\ref{14}) reduces to
\begin{align}\label{15}
S_{ij}=2[2S_1m_{[i}\bar{m}_{j]}+\bar{S}_0(l_{[i}m_{j]}+m_{[i}n_{j]})+C.C.].
\end{align}
Katkar \cite{katkar_9} transformed field equations (\ref{10}) into tetrad forms
\begin{align}\label{16}
\pi_1&=\tau_1=\lambda_1=\sigma_1=0,\nonumber\\
\rho_1&=\mu_1=2\epsilon_1=2\gamma_1=-\sqrt{2}kS_1,
\nonumber\\
\bar{\nu}_1&=\kappa_1=2\bar{\alpha}_1=2\beta_1=-\sqrt{2}kS_0.
\end{align}
We suppose Einstein-Cartan spacetime contains Weyssenhoff fluid with anisotropic matter.
\begin{align}\label{16'}
t_{ij}&=(\rho+p_t)u_iu_j-p_tg_{ij}+(p_r-p_t)v_iv_j-S_{hi;k}u^ku^hu_j.
\end{align}
The $u_i$ and $v_i$ can be chosen as,
\begin{align*}
u_i=\frac{1}{\sqrt{2}}(l_i+n_i)\hspace{0.5cm} \text{and} \hspace{0.5cm}v_i=\frac{1}{\sqrt{2}}(l_i-n_i),
\end{align*}
and so equation (\ref{16'}) written as,
\begin{align}\label{16''}
t_{ij}&=\frac{1}{2}(\rho+p_t)(l_il_j+l_in_j+n_il_j+n_in_j)-p_t(l_in_j+n_il_j-m_i\bar{m_j}-\bar{m_i}m_j) \nonumber\\
&\hspace{0.5cm}+\frac{1}{2}(p_r-p_t)(l_il_j-l_in_j-n_il_j+n_in_j) \nonumber\\
&\hspace{0.5cm}+\frac{1}{2\sqrt{2}}[\bar{S}_0\{(\bar{\epsilon}+\bar{\nu}-\kappa-\tau)+C.C\}(l_il_j+l_in_j-n_il_j-n_in_j)\nonumber\\
&\hspace{0.5cm}+\{2S_1(\pi+\nu-\bar{\kappa}-\bar{\tau})-2\bar{S}_0(\epsilon+\bar{\epsilon}+\gamma+\bar{\gamma})\}(m_il_j+m_in_j)+C.C].
\end{align}
\section{Morris-Thorne Wormhole Using Differential Forms}
Wormhole solutions must obey Einstein's field equations and have a throat connecting two asymptotically flat regions of the universe. The metric for  Morris-Thorne wormhole \cite{mt,vis95} is given by,
\begin{eqnarray}\label{17}
ds^2 = e^{2\Phi(r)}dt^{2}-\frac{dr^2}{\left(1-\frac{b(r)}{r}\right)}-r^2d\theta^2-r^2\sin^2\theta d\phi^2,
\end{eqnarray}
where $\Phi(r)$ is a red-shift function and $b(r)$ is a shape function.  There should be no event horizon in the traversable wormhole and the effect of tidal gravity forces on the traveler should be negligible. The shape function has to satisfy the following conditions in order wormhole solutions to exist: 
\begin{enumerate}[(i)]
  \item $b(r_{0})=r_{0}$ 
   \item  For all $r > r_{0}$, $\frac{b(r)-b'(r)r}{b^2(r)}>0$, this is called flare-out condition.
   \item $b'(r)< 1$  
   \item  $\frac{b(r)}{r} < 1$ for $r> r_{0}$ 
  \item  $\frac{b(r)}{r}\rightarrow$ as $r\rightarrow \infty$.
 \end{enumerate}
Note that the above requirements signify that the equality holds at the throat, $b(r) \leq r$ and $b'(r)\leq 1$ for all $r \geq r_0$.\\
We choose a set of four basis 1-forms $\theta^\alpha$ as,
\begin{align}\label{18}
\theta^1 &=\frac{1}{\sqrt{2}}\left[e^\Phi(r) dt-\left(1-\frac{b(r)}{r}\right)^{-\frac{1}{2}}dr\right], \nonumber \\
\theta^2 &=\frac{1}{\sqrt{2}}\left[e^\Phi(r) dt +\left(1-\frac{b(r)}{r}\right)^{-\frac{1}{2}}dr\right], \nonumber \\
\theta^3 &= \frac{r}{\sqrt{2}}(d\theta-i\sin\theta d\phi),\nonumber \\
\theta^4 &= \frac{r}{\sqrt{2}}(d\theta+i\sin\theta d\phi).
\end{align}
Consequently, the metric (\ref{17}) becomes
\begin{align}\label{19}
ds^2=2\theta^1\theta^2-2\theta^3\theta^4.
\end{align}
The non-vanishing covariant components of
metric tensor for the metric (\ref{19}) denoted by $\eta_{(\alpha)(\beta)}$ are given by
\begin{align}\label{20}
\eta_{12}=1=\eta_{21}, \hspace{0.5cm} \eta_{34}=-1=\eta_{43}
\end{align}
By equations (\ref{1}) and (\ref{18}), the null tetrad vectors can be chosen as
\begin{align}\label{21}
l_i&=\frac{1}{\sqrt{2}}\left[e^\Phi(r),\left(1-\frac{b(r)}{r}\right)^{-\frac{1}{2}},0,0\right],\nonumber\\
n_i&=\frac{1}{\sqrt{2}}\left[e^\Phi(r),-\left(1-\frac{b(r)}{r}\right)^{-\frac{1}{2}},0,0\right],\nonumber\\
m_i&=\frac{1}{\sqrt{2}}(0,0,-r,-ir\sin\theta),\nonumber\\
\bar{m}_i&=\frac{1}{\sqrt{2}}(0,0,-r,ir\sin\theta).
\end{align}

Using (\ref{16}) the exterior derivatives of the basis 1-forms are,
\begin{align}\label{23}
d\theta^1&=\frac{-\Phi'(r)}{\sqrt{2}}\left(1-\frac{b(r)}{r}\right)^{\frac{1}{2}}\theta^1 \wedge \theta^2-\sqrt{2}kS_0 (\theta^1 \wedge \theta^3-\theta^2 \wedge \theta^3)\nonumber\\& \hspace{0.5cm}-\sqrt{2}k\bar{S_0}(\theta^1 \wedge \theta^4-\theta^2 \wedge \theta^4)+2\sqrt{2}kS_1\theta^3 \wedge \theta^4,\nonumber\\
&=d\theta^2,\nonumber\\
d\theta^3&=-\frac{1}{\sqrt{2}r}\left(1-\frac{b(r)}{r}\right)^{\frac{1}{2}}\theta^1 \wedge \theta^3+\frac{1}{\sqrt{2}r}\left(1-\frac{b(r)}{r}\right)^{\frac{1}{2}}\theta^2 \wedge \theta^3-\frac{\cot\theta}{\sqrt{2}r}\theta^3 \wedge \theta^4, \nonumber\\
d\theta^4&=-\frac{1}{\sqrt{2}r}\left(1-\frac{b(r)}{r}\right)^{\frac{1}{2}}\theta^1 \wedge \theta^4+\frac{1}{\sqrt{2}r}\left(1-\frac{b(r)}{r}\right)^{\frac{1}{2}}\theta^2 \wedge \theta^4-\frac{\cot\theta}{\sqrt{2}r}\theta^4 \wedge \theta^3.
\end{align}
Comparing above equation with Cartan's first equation of structure (\ref{4}) the non-vanishing Ricci rotation coefficients are,
\begin{align}\label{24}
\rho&=\mu=\frac{1}{\sqrt{2}r}\left(1-\frac{b(r)}{r}\right)^{\frac{1}{2}},\nonumber\\
\alpha&=-\beta=\frac{-\cot\theta}{2\sqrt{2}r},\nonumber\\
\epsilon&=\gamma=-\frac{\Phi'(r)}{2\sqrt{2}}\left(1-\frac{b(r)}{r}\right)^{\frac{1}{2}}.
\end{align}
\indent The tetrad components of connection 1-forms are easily obtained by substituting these values in equations (\ref{6}) to get
\begin{align}\label{25}
\omega_{12}&=\frac{\Phi'(r)}{\sqrt{2}}\left(1-\frac{b(r)}{r}\right)^{\frac{1}{2}}(\theta^1+\theta^2)+\sqrt{2}kS_0 \theta^3+\sqrt{2}k\bar{S_0}\theta^4,\nonumber\\
\omega_{13}&=\sqrt{2}kS_0 \theta^1+\left[-\frac{1}{\sqrt{2}r}\left(1-\frac{b(r)}{r}\right)^{\frac{1}{2}}+\sqrt{2}kS_1\right]\theta^4,\nonumber\\
\omega_{23}&=-\sqrt{2}k{S_0} \theta^2+\left[\frac{1}{\sqrt{2}r}\left(1-\frac{b(r)}{r}\right)^{\frac{1}{2}}+\sqrt{2}kS_1\right]\theta^4,\nonumber\\
\omega_{34}&=-\sqrt{2}kS_1 (\theta^1+\theta^2)+\frac{\cot\theta}{\sqrt{2}r}(\theta^3-\theta^4).
\end{align}
Using Cartan's second equation of structure (\ref{7}) the tetrad form of curvature 2-forms are obtained as,

\begin{align*} 
\Omega_{12}&=-\left[\Phi''(r)\left(1-\frac{b(r)}{r}\right)+\frac{\Phi'(r)}{2}\left(\frac{b(r)}{r^2}-\frac{b'(r)}{r}\right)+{\Phi'(r)}^2\left(1-\frac{b(r)}{r}\right)+4k^2S_0 \bar{S_0}\right]\theta^1 \wedge \theta^2  \\
&\hspace{0.4cm}+\left[2k^2S_0S_1-kS_{0,r}\left(1-\frac{b(r)}{r}\right)^{1/2}\right]\theta^1 \wedge \theta^3+\left[-2k^2\bar{S}_0S_1-k\bar{S}_{0,r}\left(1-\frac{b(r)}{r}\right)^{1/2}\right]\theta^1 \wedge \theta^4  \\
&\hspace{0.4cm}+\left[2k^2S_0S_1+kS_{0,r}\left(1-\frac{b(r)}{r}\right)^{1/2}\right]\theta^2 \wedge \theta^3+\left[-2k^2\bar{S}_0S_1+k\bar{S}_{0,r}\left(1-\frac{b(r)}{r}\right)^{1/2}\right]\theta^2 \wedge \theta^4  \\
&\hspace{0.4cm}+\left[\frac{k}{r}(\bar{S}_0-S_0)\cot \theta-4kS_1 \left(1-\frac{b(r)}{r}\right)^{1/2}\right] \theta^3 \wedge \theta^4, 
\end{align*}

\begin{align*}
\Omega_{13}&=-\left[2kS_0 \Phi'(r) \left(1-\frac{b(r)}{r}\right)^{1/2} +kS_{0,r}\left(1-\frac{b(r)}{r}\right)^{1/2}+2k^2S_0S_1\right]\theta^1 \wedge \theta^2\\
&\hspace{0.4cm}+\left[\frac{kS_0}{r}\cot\theta-2k^2S_0^2\right]\theta^1 \wedge \theta^3
+\left[\frac{1}{4r}\left(\frac{b(r)}{r^2}-\frac{b'(r)}{r}\right)-kS_{1,r}\left(1-\frac{b(r)}{r}\right)^{1/2}\right.\\
&\hspace{0.4cm}\left.+kS_1\Phi'(r)\left(1-\frac{b(r)}{r}\right)^{1/2} -\frac{\Phi'(r)}{2r}\left(1-\frac{b(r)}{r}\right)
-\frac{kS_0}{r}\cot\theta-2k^2S_1^2-2k^2S_0\bar{S_0}\right]\theta^1 \wedge \theta^4\\
&\hspace{0.4cm}+\left[\frac{-1}{4r}\left(\frac{b(r)}{r^2}-\frac{b'(r)}{r}\right)+kS_{1,r}\left(1-\frac{b(r)}{r}\right)^{1/2}
 kS_1\Phi'(r)\left(1-\frac{b(r)}{r}\right)^{1/2}-\frac{\Phi'(r)}{2r}\left(1-\frac{b(r)}{r}\right)\right.\\
 &\hspace{0.4cm}\left.+\frac{2kS_1}{r}\left(1-\frac{b(r)}{r}\right)^{1/2}-2k^2S_1^2\right]\theta^2 \wedge \theta^4 +\left[\frac{-kS_0}{r}\left(1-\frac{b(r)}{r}\right)^{1/2}+2k^2S_0S_1\right]\theta^3 \wedge \theta^4,
\end{align*}

\begin{align*}
\Omega_{23}&=-\left[2k^2S_0S_1-kS_{0,r}\left(1-\frac{b(r)}{r}\right)^{1/2}-2kS_0\Phi'(r)\left(1-\frac{b(r)}{r}\right)^{1/2}\right]\theta^1 \wedge \theta^2\nonumber\\
&\hspace{0.4cm}-\left[kS_{1,r}\left(1-\frac{b(r)}{r}\right)^{1/2}+\frac{1}{4r}\left(\frac{b(r)}{r^2}-\frac{b'(r)}{r}\right)+kS_1\Phi'(r)\left(1-\frac{b(r)}{r}\right)^{1/2}+\frac{\Phi'(r)}{2r}\left(1-\frac{b(r)}{r}\right)\right. \nonumber\\
&\hspace{0.4cm}+\left.\frac{2kS_1}{r}\left(1-\frac{b(r)}{r}\right)^{1/2}+2k^2S_1^2\right]\theta^1 \wedge \theta^4-\left[\frac{kS_0}{r}\cot\theta+2k^2S_0^2\right]\theta^2 \wedge \theta^3 \nonumber\\
&\hspace{0.4cm}+\left[kS_{1,r}\left(1-\frac{b(r)}{r}\right)^{1/2}+\frac{1}{4r}\left(\frac{b(r)}{r^2}-\frac{b'(r)}{r}\right)-\frac{\Phi'(r)}{2r}\left(1-\frac{b(r)}{r}\right)-kS_1\Phi'(r)\left(1-\frac{b(r)}{r}\right)^{1/2}\right.\nonumber\\
&\hspace{0.4cm}+\left.\frac{kS_0}{r}\cot\theta-\frac{2kS_1}{r}\left(1-\frac{b(r)}{r}\right)^{1/2}-2k^2S_1^2-2k^2S_0\bar{S_0}\right]\theta^2 \wedge \theta^4\nonumber\\
&\hspace{0.4cm}-\left[\frac{kS_0}{r}\left(1-\frac{b(r)}{r}\right)^{1/2}+2k^2S_0S_1\right]\theta^3 \wedge \theta^4,
\end{align*}
\begin{align}\label{ss}
\Omega_{34}&=\left[2kS_{1,r}\left(1-\frac{b(r)}{r}\right)^{1/2}+2kS_{1}\Phi'(r)\left(1-\frac{b(r)}{r}\right)^{1/2}\right]\theta^1 \wedge \theta^2\nonumber\\
&+\left[\frac{2kS_1}{r}\cot\theta-\frac{kS_0}{r}\left(1-\frac{b(r)}{r}\right)^{1/2}+2k^2S_0S_1\right]\theta^1 \wedge \theta^3\nonumber\\
&+\left[\frac{2kS_1}{r}\cot\theta+\frac{k \bar{S_0}}{r}\left(1-\frac{b(r)}{r}\right)^{1/2}+2k^2\bar{S_0}S_1\right]\theta^1 \wedge \theta^4\nonumber\\
&+\left[\frac{2kS_1}{r}\cot\theta-\frac{kS_0}{r}\left(1-\frac{b(r)}{r}\right)^{1/2}-2k^2S_0S_1\right]\theta^2 \wedge \theta^3\nonumber\\
&+\left[\frac{2kS_1}{r}\cot\theta+\frac{k \bar{S_0}}{r}\left(1-\frac{b(r)}{r}\right)^{1/2}-2k^2\bar{S_0}S_1\right]\theta^2 \wedge \theta^4+\left[\frac{b(r)}{r^3}-4k^2S_1^2\right]\theta^3 \wedge \theta^4.
\end{align}

The relation of curvature 2-forms and Riemann-Christoffel curvature tensor given in equation (\ref{8}) and hence by equations (\ref{8}) and (\ref{ss}) the independent tetrad component of Riemann tensor are obtained as,

 \begin{align*}
R_{1212}&=-\left[\Phi''(r)\left(1-\frac{b(r)}{r}\right)+\frac{\Phi'(r)}{2}\left(\frac{b(r)}{r^2}-\frac{b'(r)}{r}\right)+{\Phi'(r)}^2\left(1-\frac{b(r)}{r}\right)+4k^2S_0 \bar{S_0}\right],\\
R_{1312}&=\left[2k^2S_0S_1-kS_{0,r}\left(1-\frac{b(r)}{r}\right)^{1/2}\right],\\
R_{2312}&=\left[2k^2S_0S_1+kS_{0,r}\left(1-\frac{b(r)}{r}\right)^{1/2}\right],\\
R_{3412}&=\left[\frac{k}{r}(\bar{S}_0-S_0)\cot \theta-\frac{4kS_1}{r} \left(1-\frac{b(r)}{r}\right)^{1/2}\right],\\
R_{1213}&=\left[-2kS_0 \Phi' \left(1-\frac{b(r)}{r}\right)^{1/2} -kS_{0,r}\left(1-\frac{b(r)}{r}\right)^{1/2}-2k^2S_0S_1\right],\\
R_{1313}&=\left[\frac{kS_0}{r}\cot\theta-2k^2S_0^2\right],\\
R_{1413}&=\left[\frac{1}{4r}\left(\frac{b(r)}{r^2}-\frac{b'(r)}{r}\right)-kS_{1,r}\left(1-\frac{b(r)}{r}\right)^{1/2}+kS_1\Phi'(r)\left(1-\frac{b(r)}{r}\right)^{1/2}\right. \\
&\hspace{0.5cm}-\left.\frac{\Phi'(r)}{2r}\left(1-\frac{b(r)}{r}\right)-\frac{kS_0}{r}\cot\theta-2k^2S_1^2-2k^2S_0\bar{S_0}\right],\\
R_{2313}&=0,
\end{align*}
\begin{align}\label{27}
R_{2413}&=\left[\frac{-1}{4r}\left(\frac{b(r)}{r^2}-\frac{b'(r)}{r}\right)+kS_{1,r}\left(1-\frac{b(r)}{r}\right)^{1/2}+kS_1\Phi'\left(1-\frac{b(r)}{r}\right)^{1/2}\right.\nonumber\\
&\hspace{0.5cm}-\left.\frac{\Phi'(r)}{2r}\left(1-\frac{b(r)}{r}\right)+\frac{2kS_1}{r}\left(1-\frac{b(r)}{r}\right)^{1/2}-2k^2S_1^2\right],\nonumber\\
R_{3413}&=\left[\frac{-kS_0}{r}\left(1-\frac{b(r)}{r}\right)^{1/2}+2k^2S_0S_1\right],\nonumber\\
R_{1223}&=\left[kS_{0,r}\left(1-\frac{b(r)}{r}\right)^{1/2}+2kS_0\Phi'(r)\left(1-\frac{b(r)}{r}\right)^{1/2}-2k^2S_0S_1\right],\nonumber\\
R_{1323}&=0,\nonumber\\
R_{1423}&=-\left[kS_{1,r}\left(1-\frac{b(r)}{r}\right)^{1/2}+\frac{1}{4r}\left(\frac{b(r)}{r^2}-\frac{b'(r)}{r}\right)+kS_1\Phi'(r)\left(1-\frac{b(r)}{r}\right)^{1/2}\right. \nonumber \\
&\hspace{0.5cm}+\left.\frac{\Phi'(r)}{2r}\left(1-\frac{b(r)}{r}\right)+\frac{2kS_1}{r}\left(1-\frac{b(r)}{r}\right)^{1/2}+2k^2S_1^2\right],\nonumber\\
R_{2323}&=-\left[\frac{kS_0}{r}\cot\theta+2k^2S_0^2\right],\nonumber\\
R_{2423}&=\left[kS_{1,r}\left(1-\frac{b(r)}{r}\right)^{1/2}+\frac{1}{4r}\left(\frac{b(r)}{r^2}-\frac{b'(r)}{r}\right)-\frac{\Phi'(r)}{2r}\left(1-\frac{b(r)}{r}\right)\right. \nonumber\\
&\hspace{0.5cm}-\left.kS_1\Phi'\left(1-\frac{b(r)}{r}\right)^{1/2}+\frac{kS_0}{r}\cot\theta-\frac{2kS_1}{r}\left(1-\frac{b(r)}{r}\right)^{1/2}-2k^2S_1^2-2k^2S_0\bar{S_0}\right], \nonumber\\
R_{3423}&=-\left[\frac{kS_0}{r}\left(1-\frac{b(r)}{r}\right)^{1/2}+2k^2S_0S_1\right],\nonumber\\
R_{1234}&=\left[2kS_{1,r}\left(1-\frac{b(r)}{r}\right)^{1/2}+2kS_{1}\Phi'\left(1-\frac{b(r)}{r}\right)^{1/2}\right],\nonumber\\
R_{1334}&=\left[\frac{2kS_1}{r}\cot\theta-\frac{kS_0}{r}\left(1-\frac{b(r)}{r}\right)^{1/2}+2k^2S_0S_1\right],\nonumber\\
R_{1434}&=\left[\frac{2kS_1}{r}\cot\theta+\frac{k \bar{S_0}}{r}\left(1-\frac{b(r)}{r}\right)^{1/2}+2k^2\bar{S_0}S_1\right],\nonumber\\
R_{2334}&=\left[\frac{2kS_1}{r}\cot\theta-\frac{kS_0}{r}\left(1-\frac{b(r)}{r}\right)^{1/2}-2k^2S_0S_1\right],\nonumber\\
R_{2434}&=\left[\frac{2kS_1}{r}\cot\theta+\frac{k \bar{S_0}}{r}\left(1-\frac{b(r)}{r}\right)^{1/2}-2k^2\bar{S_0}S_1\right],\nonumber\\
R_{3434}&=\left[\frac{b(r)}{r^3}-4k^2S_1^2\right]
\end{align}
The tetrad components of Ricci tensor and Ricci scalar are expressed as,
\begin{align}\label{28}
R_{11}&=\frac{1}{2r}\left(\frac{b(r)}{r^2}-\frac{b'(r)}{r}\right)-\frac{\Phi'(r)}{r}\left(1-\frac{b(r)}{r}\right)-\frac{k}{r}\cot\theta(S_0+\bar{S_0})-4k^2S_1^2-4k^2S_0\bar{S_0},\nonumber\\
R_{12}&=-\left[\left(\Phi''(r)+{\Phi'(r)}^2+\frac{\Phi'(r)}{r}\right)\left(1-\frac{b(r)}{r}\right)+\left(\frac{\Phi'(r)}{2}+\frac{1}{2r}\right)\left(\frac{b(r)}{r^2}-\frac{b'(r)}{r}\right)+4k^2S_1^2+4k^2S_0\bar{S_0}\right]\nonumber\\
&=R_{21},\nonumber\\
R_{13}&=\frac{2kS_1}{r}\cot\theta-\left[\frac{kS_0}{r}+kS_{0,r}+2kS_0\Phi'(r)\right]\left(1-\frac{b(r)}{r}\right)^{1/2},\nonumber\\
R_{22}&=\frac{1}{2r}\left(\frac{b(r)}{r^2}-\frac{b'(r)}{r}\right)-\frac{\Phi'(r)}{r}\left(1-\frac{b(r)}{r}\right)+\frac{k}{r}\cot\theta(S_0+\bar{S_0})-4k^2S_1^2-4k^2S_0\bar{S_0},\nonumber\\
R_{23}&=\frac{2kS_1}{r}\cot\theta-k\left[2\Phi'(r)S_0+\frac{S_0}{r}+S_{0,r}\right]\left(1-\frac{b(r)}{r}\right)^{1/2},\nonumber\\
R_{31}&=4k^2S_0S_1-\left[\frac{kS_0}{r}+kS_{0,r}\right]\left(1-\frac{b(r)}{r}\right)^{1/2}, \nonumber\\
R_{32}&=-\left[\left(kS_{0,r}+\frac{kS_0}{r}\right)\left(1-\frac{b(r)}{r}\right)^{1/2}+4k^2S_0S_1\right],\nonumber\\
R_{33}&=0,\nonumber\\
R_{34}&=\frac{1}{2r}\left(\frac{b(r)}{r^2}-\frac{b'(r)}{r}\right)+\frac{\Phi'(r)}{r}\left(1-\frac{b(r)}{r}\right)+\frac{b(r)}{r^3},
\end{align}
and
\begin{align}
R&=-2\left[\Phi''(r)+{\Phi'(r)}^2+\frac{2\Phi'(r)}{r}\right]\left(1-\frac{b(r)}{r}\right)-\left(\Phi'(r)+\frac{2}{r}\right)\left(\frac{b(r)}{r^2}-\frac{b'(r)}{r}\right)\nonumber\\
&\hspace{0.5cm}-\frac{2b(r)}{r^3}-8k^2S_0\bar{S_0}-8k^2S_1^2.
\end{align}
\indent In the tetrad form Energy momentum tensor can be obtained by,
\begin{align*}
t_{\alpha\beta}=t_{ij}e^i\hspace{0.1mm}_{(\alpha)}e^j\hspace{0.1mm}_{(\beta)}
\end{align*}
Hence by equation (\ref{16''}) we get,
\begin{align}\label{29}
t_{11}&=t_{22}=\frac{1}{2}(\rho+p_r),\hspace{0.5cm}
t_{12}=t_{21}=\frac{1}{2}(\rho-p_r),\nonumber\\
t_{34}&=t_{43}=p_t, \hspace{0.5cm} t_{13}=t_{23}=t_{33}=0,\nonumber\\
t_{31}&=t_{32}=-S_0\Phi'(r)\left(1-\frac{b(r)}{r}\right)^{1/2}.
\end{align}
\subsection{Field Equations}
Field equations are expressed as,
\begin{align}\label{30}
&R_{ij}-\frac{1}{2}\eta_{ij}R=-kt_{ij},\nonumber\\
&\frac{1}{2r}\left(\frac{b(r)}{r^2}-\frac{b'(r)}{r}\right)-\frac{\Phi'(r)}{r}\left(1-\frac{b(r)}{r}\right)-\frac{k}{r}\cot\theta(S_0+\bar{S_0})-4k^2S_1^2-4k^2S_0\bar{S_0}=\frac{-k}{2}(\rho+p_r),\nonumber\\
&\frac{\Phi'(r)}{r}\left(1-\frac{b(r)}{r}\right)+\frac{1}{2r}\left(\frac{b(r)}{r^2}-\frac{b'(r)}{r}\right)+\frac{b(r)}{r^3}=\frac{-k}{2}(\rho-p_r),\nonumber\\
&\frac{2kS_1}{r}\cot\theta-\left[\frac{kS_0}{r}+kS_{0,r}+kS_0\Phi'(r)\right]\left(1-\frac{b(r)}{r}\right)^{1/2}=0,\nonumber\\
&\frac{1}{2r}\left(\frac{b(r)}{r^2}-\frac{b'(r)}{r}\right)-\frac{\Phi'(r)}{r}\left(1-\frac{b(r)}{r}\right)+\frac{k}{r}\cot\theta(S_0+\bar{S_0})-4k^2S_1^2-4k^2S_0\bar{S_0}=\frac{-k}{2}(\rho+p_r),\nonumber\\
&\frac{2kS_1}{r}\cot\theta-\left[\frac{kS_0}{r}+kS_{0,r}+2kS_0\Phi'(r)\right]\left(1-\frac{b(r)}{r}\right)^{1/2}=0,\nonumber\\
&4k^2S_0S_1-\left[kS_0\left(\frac{1}{r}+1\right)+kS_{0,r}\right]\left(1-\frac{b(r)}{r}\right)^{1/2}=0,\nonumber\\
&4k^2S_0S_1+\left[kS_0\Phi'(r)+S_{0,r}\right]\left(1-\frac{b(r)}{r}\right)^{1/2}=0, \nonumber\\
&\left(\frac{\Phi'(r)}{2}+\frac{1}{2r}\right)\left(\frac{b(r)}{r^2}-\frac{b'(r)}{r}\right)+\left(\Phi''(r)+{\Phi'(r)}^2+\frac{\Phi'(r)}{r}\right)\left(1-\frac{b(r)}{r}\right)\nonumber\\
&+4k^2S_0\bar{S}_0+4k^2S_1^2=kp_t.
\end{align}

In equations (\ref{30}) one can see that they are consistent only if $S_0=0$ and so they reduce to
\begin{align}\label{31}
&\frac{1}{2r}\left(\frac{b(r)}{r^2}-\frac{b'(r)}{r}\right)-\frac{\Phi'(r)}{r}\left(1-\frac{b(r)}{r}\right)-4k^2S_1^2=\frac{-k}{2}(\rho+p_r), \nonumber \\
&\frac{\Phi'(r)}{r}\left(1-\frac{b(r)}{r}\right)+\frac{1}{2r}\left(\frac{b(r)}{r^2}-\frac{b'(r)}{r}\right)+\frac{b(r)}{r^3}=\frac{-k}{2}(\rho-p_r), \nonumber\\
&\left(\frac{\Phi'(r)}{2}+\frac{1}{2r}\right)\left(\frac{b(r)}{r^2}-\frac{b'(r)}{r}\right)+\left(\Phi''(r)+{\Phi'(r)}^2+\frac{\Phi'(r)}{r}\right)\left(1-\frac{b(r)}{r}\right)\nonumber\\
&+4k^2S_1^2=kp_t.
\end{align}
By equations (\ref{31}) density and pressures are written as,
\begin{align}
\label{ef_rho}
k\rho&=4k^2S_1^2-\frac{1}{r}\left(\frac{b(r)}{r^2}-\frac{b'(r)}{r}\right)-\frac{b(r)}{r^3},  \\
\label{ef_pr}
kp_r&=4k^2S_1^2+\frac{2\Phi'(r)}{r}\left(1-\frac{b(r)}{r}\right)+\frac{b(r)}{r^3}, \\
\label{ef_pt}
kp_t&=\left(\frac{\Phi'(r)}{2}+\frac{1}{2r}\right)\left(\frac{b(r)}{r^2}-\frac{b'(r)}{r}\right)+\left(\Phi''(r)+{\Phi'(r)}^2+\frac{\Phi'(r)}{r}\right)\left(1-\frac{b(r)}{r}\right)\nonumber\\
&\hspace{0.5cm}+4k^2S_1^2.
\end{align}
However, in the above expressions there is a term of spin density $S_1$ for that we used the law of conservation of energy-momentum tensor from which we obtained the equation,
\begin{align}\label{33}
\Phi'(r)(\rho+p_r)+p'_r+\frac{2}{r}(p_r-p_t)-4[2\Phi'(r)S_1^2+(S_1^2)']=0
\end{align}
In equation (\ref{33}) the first three terms correspond to the matter part and the last term corresponds to the spin density part. The conservation of the spin part can be written independently as,
\begin{align}\label{34}
2\Phi'(r)S_1^2+(S_1^2)'=0
\end{align}
The solution to the above equation is
\begin{equation}\label{35}
S_1^2=Ce^{-2\Phi(r)}
\end{equation}
where $C$ is a positive integration constant.

%\section{Geometric embedding of wormhole}
%In this section, we shall revisit the embedding of wormhole geometry with the help of two adopted distinct shape functions. Sahoo et al. \cite{sahoo1_2023,sahoo2_2023}, seminal work provided  shape function  $b(r)=r_0\sqrt{\frac{r}{r_0}}$ (SF-1) and $b(r)=re^{1-\frac{r}{r_0}}$ (SF-2) with the ECs constraints for wormhole in $f (R, T)$ and $f(R,L_{m})$ gravity theories. Here, SF-1 is algebraic and SF-2 is the exponential form of a specific shape function and $r_{0}=1$. For this particular choice of shape functions, we choose one red-shift function and it is expressed as : $\Phi(r)= \frac{r_{0}}{r^n}$.

\section{Physical analysis}
An appropriate shape function $b(r)$ and red-shift function $\Phi(r)$  are needed to investigate the behavior of the matter distribution supporting the wormhole geometry. In the present investigation, we employed a particular choice of shape function (SF) and the red-shift function to perform physical analysis. The  shape function is expressed as: $b(r)=re^{1-\frac{r}{r_0}}$ \cite{sahoo1_2023}. The variable $r$ has values ranging from $r_{0}$ to infinity, where $r_{0}$ represents the minimum radius of the wormhole's throat.

%\begin{figure}[H]
%    \centering

   % \begin{subfigure}[b]{0.45\textwidth}
    %    \centering
     %   \includegraphics[width=\textwidth]{figure/br_1.png}
      %  \caption{SF-1 : $b(r)=r_0\sqrt{\frac{r}{r_0}}$ }
      %  \label{fig:sub_sf1}
   % \end{subfigure}
  % \begin{subfigure}[b]{0.45\textwidth}
    %    \centering
   %     \includegraphics[width=\textwidth]{figure/flare_sf1.png}
    %    \caption{Flare-out condition:  $\frac{b(r)-b'(r)r}{b^2(r)}>0$}
     %   \label{fig:sub_sf11}
   % \end{subfigure}

    %\caption{\textbf{Left Panel (a)} Profile of shape functions $b(r)$ and the red-shift function ($\Phi(r)=\frac{r_{0}}{r^n}$, with $r_{0}=1$, $n=3$), \textbf{Right Panel (b)} The flare-out condition against the radial coordinate. The vertical grey band represents the throat radius region, $r_{0}\leq 1$. }
    %\label{fig:br_sf1}
%\end{figure}

\begin{figure}[H]
    \centering

    \begin{subfigure}[b]{0.45\textwidth}
        \centering
        \includegraphics[width=\textwidth]{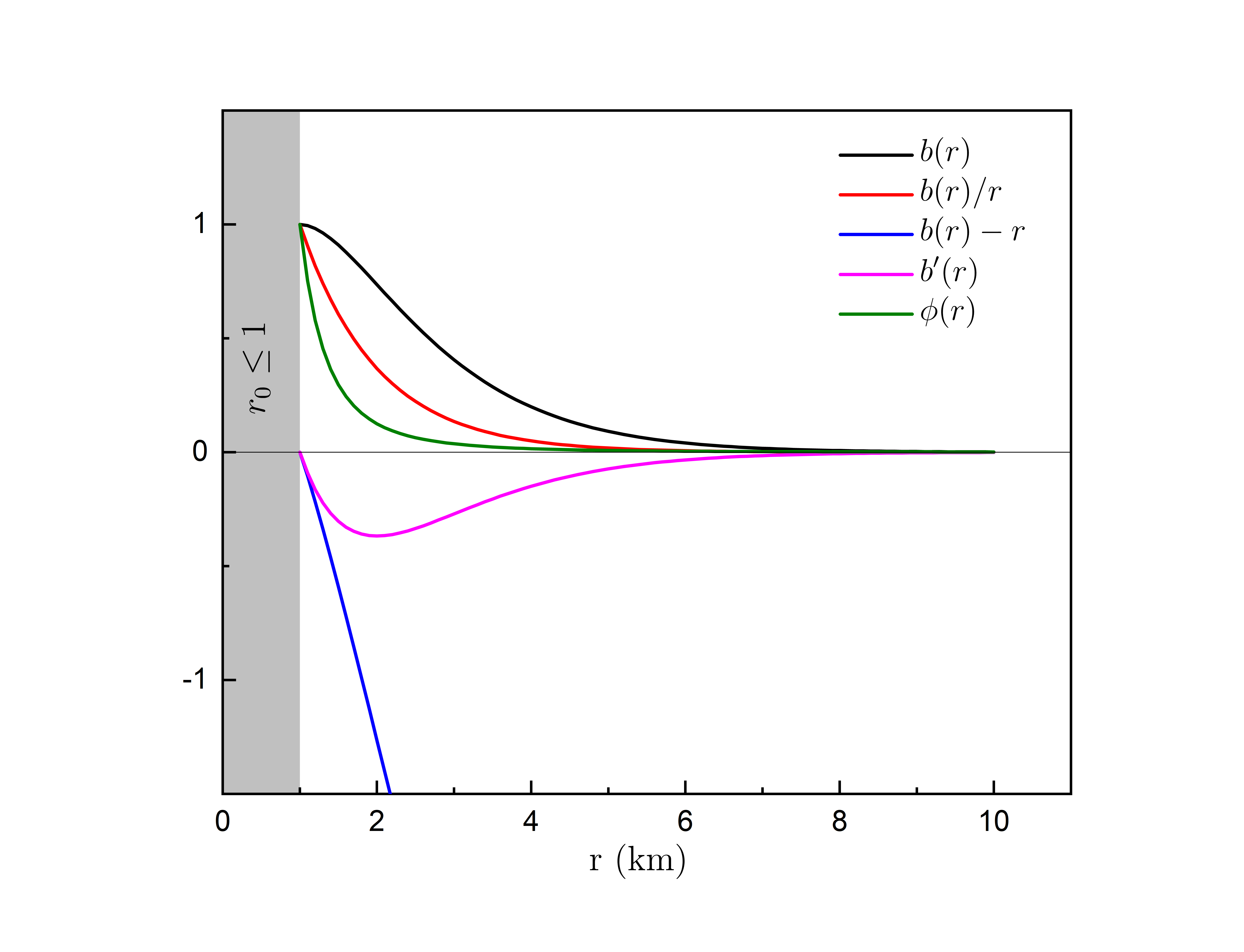}
        \caption{SF : $b(r)=re^{1-\frac{r}{r_0}}$ }
        \label{fig:sub_sf2}
    \end{subfigure}
   \begin{subfigure}[b]{0.45\textwidth}
        \centering
        \includegraphics[width=\textwidth]{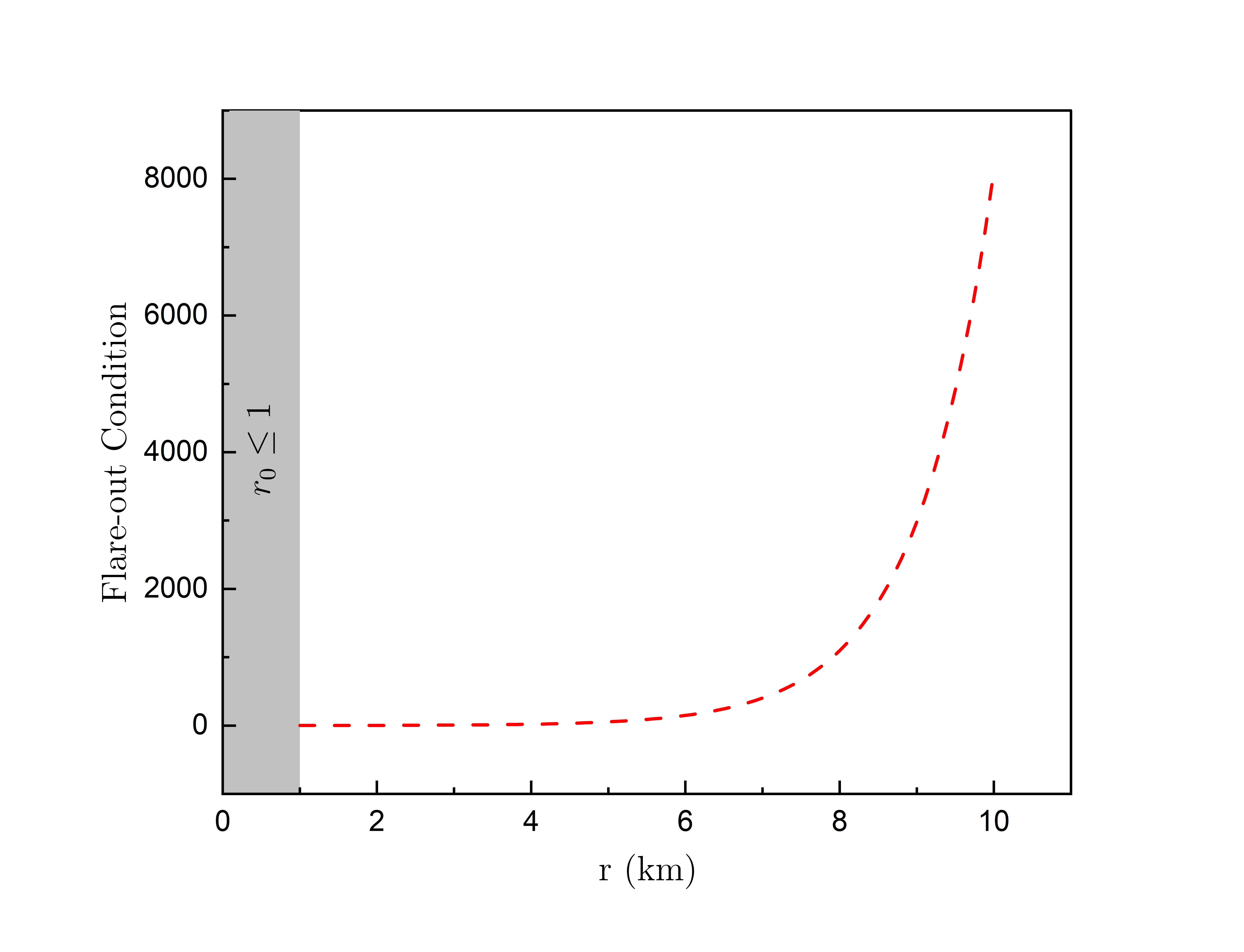}
        \caption{Flare-out condition:  $\frac{b(r)-b'(r)r}{b^2(r)}>0$}
        \label{fig:sub_sf22}
    \end{subfigure}

    \caption{\textbf{Left Panel (a)} Profile of shape functions $b(r)$ and the red-shift function ($\Phi(r)=\frac{r_{0}}{r^n}$, with $r_{0}=1$, $n=3$), \textbf{Right Panel (b)} The flare-out condition against the radial coordinate. The vertical grey band represent throat radius region, $r_{0}\leq 1$ }
    \label{fig:br_sf2}
\end{figure}

Employing this particular choice of SF, we investigate different cases. The features of the shape function and the trend of the red-shift function are shown in Fig. \ref{fig:br_sf2}. It is evident that the SF fairly meets the requirement for a  wormhole, as it can be observed that $b(r)<1$, $b'(r)< 1$, and $\frac{b(r)}{r}\rightarrow 0$ as $r\rightarrow \infty$ hold. However, in Fig.\ref{fig:sub_sf2} it can be seen that the $b '(r)<1$ criterion for the particular choice of shape function does not satisfy the geometrical requirement of a wormhole. Finally, it can be seen that all $r$ satisfy the flare-out criterion in  Fig.\ref{fig:sub_sf22}. Therefore, we infer the conclusion that the Einstein-Cartan theory and the models we have at present meet the essential criteria to characterize a traversable wormhole.

The consequences of the Einstein-Cartan theory on the hydrodynamic balance at the throat and its surroundings for the Morris-Thorne model with a non-zero red-shift function are addressed in the next part.

\subsection{ Energy Conditions
}

Energy conditions are critical for studying the behavior of matter inside the wormhole. In GR, there are some  fundamental energy requirements as follows:

\begin{itemize}
    \item Null Energy Condition (NEC) : Both $\rho+p_{r}$ and $\rho+p_{t}$ are non-negative. It depicts gravity's attractive nature. 
    \item Weak Energy Condition (WEC) : Non-negative energy density requires that $\rho+p_{r}$ and $\rho+p_{t}$ are both non-negative, besides this the energy density needs to be positive. Thus, WEC demands positive energy density besides NEC. 
    \item  Strong Energy Condition (SEC) : For non-negative $\rho + p_{r}\geq 0$, $\rho + p_{t}\geq 0$, $\rho +  p_{r} + p_t \geq 0$. It originates from the spherically symmetrical metric and the attracting character of gravity.
    \item Dominant Energy Condition (DEC) : It is given by $\rho-|p_{r}|$ and $\rho -|p_{t}|$ have to be non-negative for there to be non-negative energy density. In addition, which quantifies the velocity at which energy flows at the speed of light.
\end{itemize}

Now, we will examine the above energy conditions for a particular choice of shape function and red-shift function for the wormhole.

\subsubsection*{\textbf{Case :}  $b(r)=re^{1-\frac{r}{r_0}}$,  $\Phi(r)=\frac{r_0}{r^n}$} 

To begin with positive real number $n$, red-shift function $\Phi(r)= \frac{r_{0}}{r^n}$  and the shape function $b(r)=re^{1-\frac{r}{r_0}}$, where $n$ is arbitrary positive real number. The analysis of this kind of shape function for constant red-shift has been found in \cite{sahoo1_2023}.

In this instance, in order to solve the field equations  (\ref{ef_rho})-(\ref{ef_pt}), we took into account the exponential forms of the shape function. Following the solution of these equations, the terms for energy density, radial pressure, and tangential pressure are as follows:

\begin{align}
\label{rho_sf2}
\rho&=\frac{1}{8\pi}\left[256\pi^2 C e^{-2\frac{r_0}{r^n}}-\frac{1}{rr_0}e^{1-\frac{r}{r_0}}-\frac{1}{r^2} e^{1-\frac{r}{r_0}}\right]\\
\label{pr_sf2}
p_r&=\frac{1}{8\pi}\left[256\pi^2 Ce^{-2\frac{r_0}{r^n}}-\frac{2nr_0}{r^{n+2}}(1-e^{1-\frac{r}{r_0}})+\frac{1}{r^2}e^{1-\frac{r}{r_0}}\right]\\
\label{pt_sf2}
p_t&=\frac{1}{8\pi}\left[256\pi^2 Ce^{-2\frac{r_0}{r^n}}+\left(\frac{1}{2rr_0}-\frac{nr_0}{2r_0r^{n+1}}\right)e^{1-\frac{r}{r_0}} \right. \nonumber\\
& \quad \left. +\left(\frac{n(n+1)r_0}{r^{n+2}}+\frac{n^2r^2_0}{r^{2(n+1)}}-\frac{nr_0}{r^{n+2}}\right)
 (1-e^{1-\frac{r}{r_0}}) \right]
 \end{align}
 The pressure anisotropy is denoted by $\Delta$ and it is given by
\begin{align}
\Delta &= p_{t}-p_{r}
\end{align}

\begin{figure}[H]
    \centering
    \includegraphics[scale=0.3]{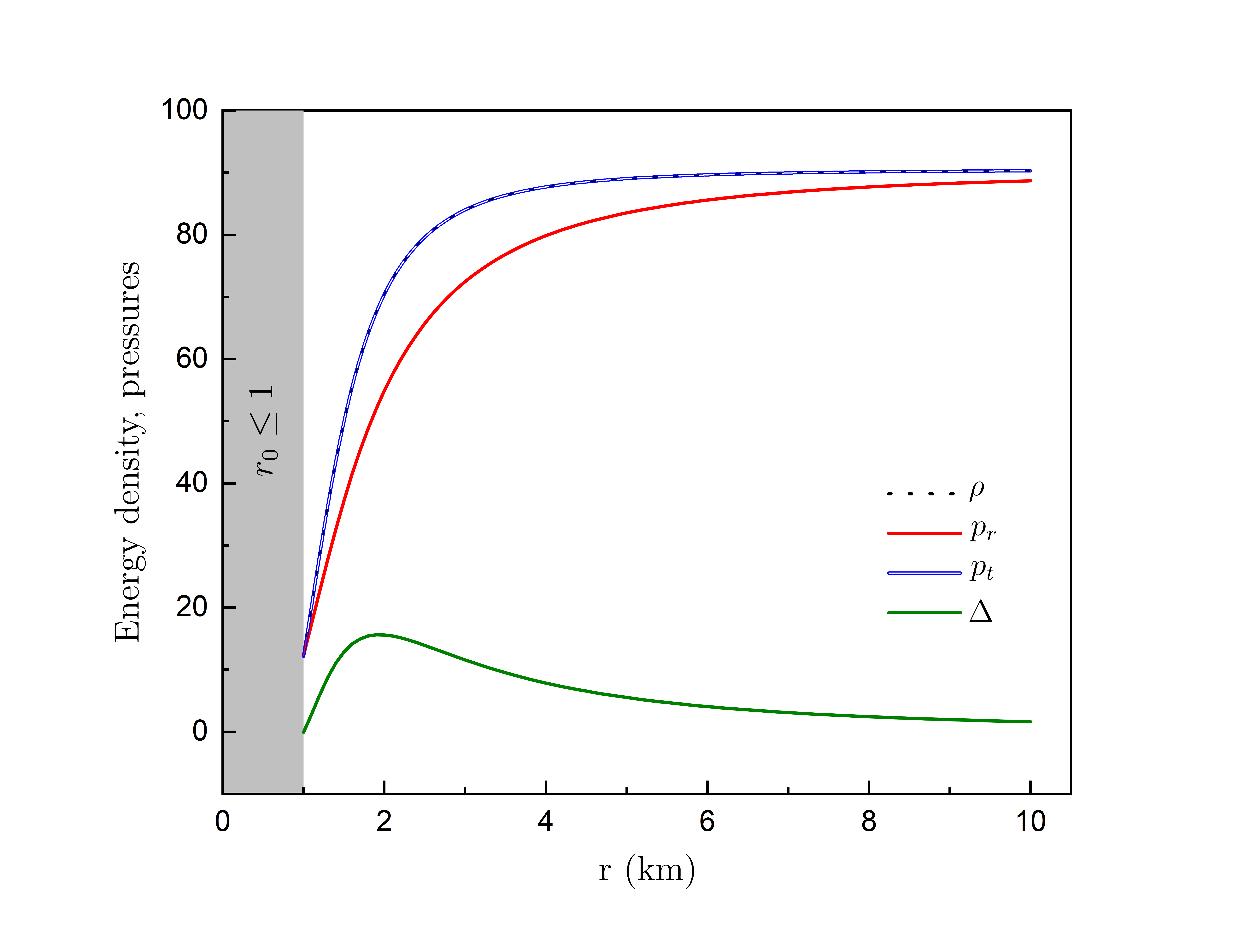}
    \caption{Variation of $\rho$, $p_{r}$, $p_{t}$ and $\Delta$ with radial coordinate $r$ for particular choice of SF with $n=3$ , $r_{0}=1$ and $C=0.9$. The vertical grey band represent throat radius region, $r_{0}\leq 1$.}
    \label{fig:rho_sf2}
\end{figure}

We show the energy density and pressure profile for the SF in Fig. \ref{fig:rho_sf2}. It can be observed that the energies $\rho$, $p_{r}$ $p_t$, and $\Delta$ are displaying a positive magnitude right above the throat.  In this study, we incorporate an anisotropic configuration, which offers the plausibility of such exotic matter. Even more, far from the throat the anisotropy $\Delta$  is saturated, i.e, $\Delta \rightarrow 0$ for large distances. Moreover, as can be seen $\rho $, $p_{r}$ and $p_{t}$ are finite for all $ r \in [r_{0},\infty ]$. The results of Fig.\ref{fig:rho_sf2} show that  $\rho$ and $p_{t}$ yield the same sort of curve.

\begin{figure}[H]
    \centering
    \includegraphics[scale=0.3]{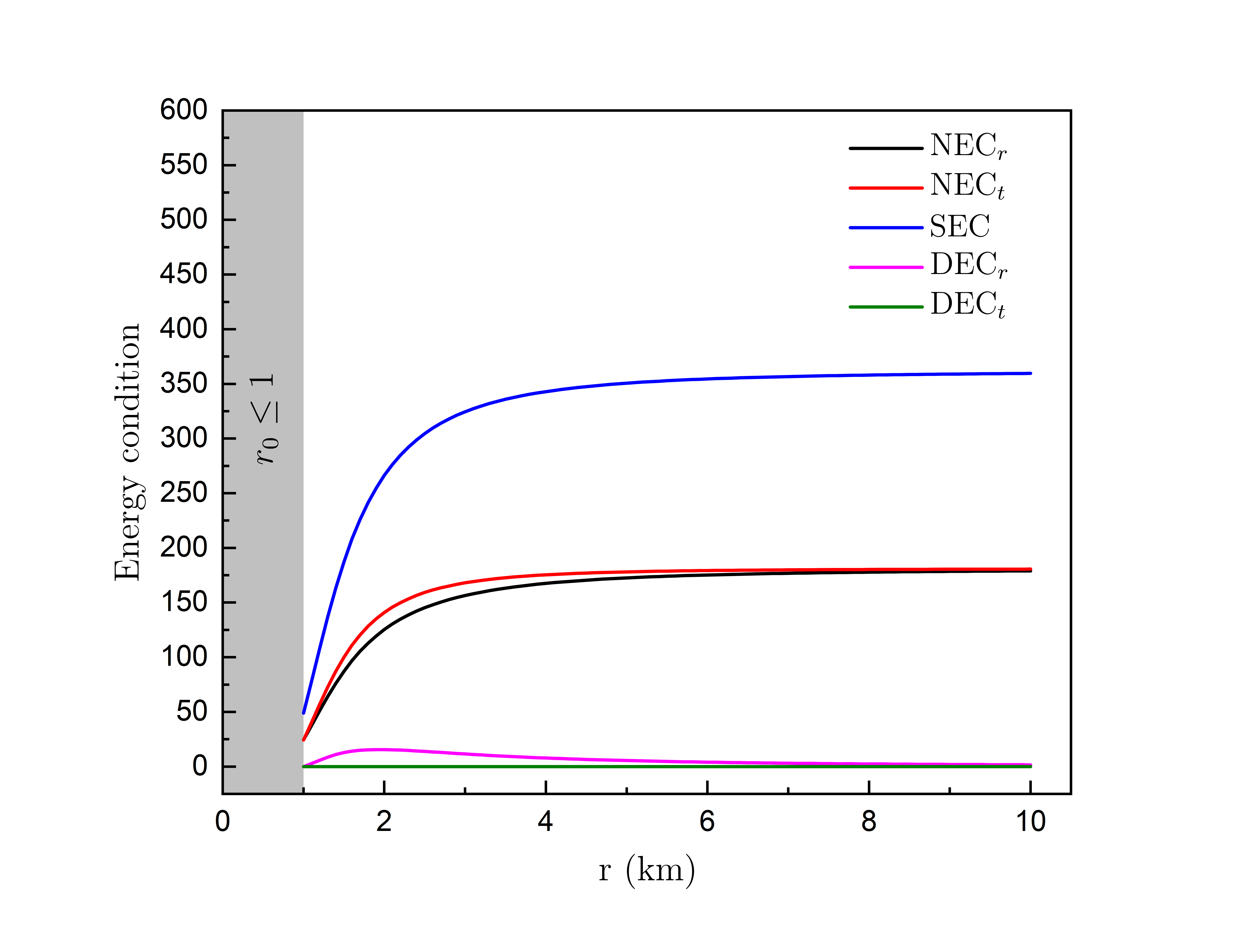}
    \caption{Variation of energy conditions, NEC, SEC and DEC against radial coordinate $r$  with $n=3$ , $r_{0}=1$ and $C=0.9$. The vertical grey band represent throat radius region, $r_{0}\leq 1$.}
    \label{fig:EC_sf2}
\end{figure}

In Fig. \ref{fig:EC_sf2}, we present the energy condition profile. It is important to note that, all the energy conditions are strictly positive throughout the entire spacetime.

\section{Stability analysis}

In this section, we investigate the stability of a wormhole within the framework of the Einstein-Cartan theory. By incorporating the spin of particles into the theory, we aim to explore how this additional geometric quantities affect the stability of the wormhole. To analyze its stability, we utilize an equilibrium condition derived from the Tolman-Oppenheimer-Volkov (TOV) equation \cite{ponce_1993,rahaman_2014}. This equation provides insights into the equilibrium state of a self-gravitating system, enabling us to assess the stability of the wormhole under consideration.\\
\noindent The four terms in Eq.(\ref{33}), which are defined as follows, are used to calculate the equilibrium state of a structure:\\
\begin{enumerate}[(i)]
\item The gravitational force
\begin{eqnarray}
F_{g}=\Phi ' (\rho + p_{r}),
\end{eqnarray}
\item The hydrostatic force 
 \begin{eqnarray}
     F_{h}=\frac{dp_{r}}{dr},
 \end{eqnarray}
\item The anisotropy force
 \begin{eqnarray}
     F_{a}= \frac{2 (p_{r}-p_{t})}{r},
\end{eqnarray}
\item The spin force
\begin{eqnarray}
    F_{s}=- 4[2\Phi'S_1^2+(S_1^2)'],
\end{eqnarray}
\end{enumerate}
due to the anisotropic pressure and spin force in a Morris-Thorne wormhole. Eq.(\ref{33}) then yields the following equilibrium condition:

\begin{eqnarray}
    F_{g}+ F_{h}+F_{a}+F_{s}=0.
\end{eqnarray}

\begin{figure}[H]
    \centering
    \includegraphics[scale=0.3]{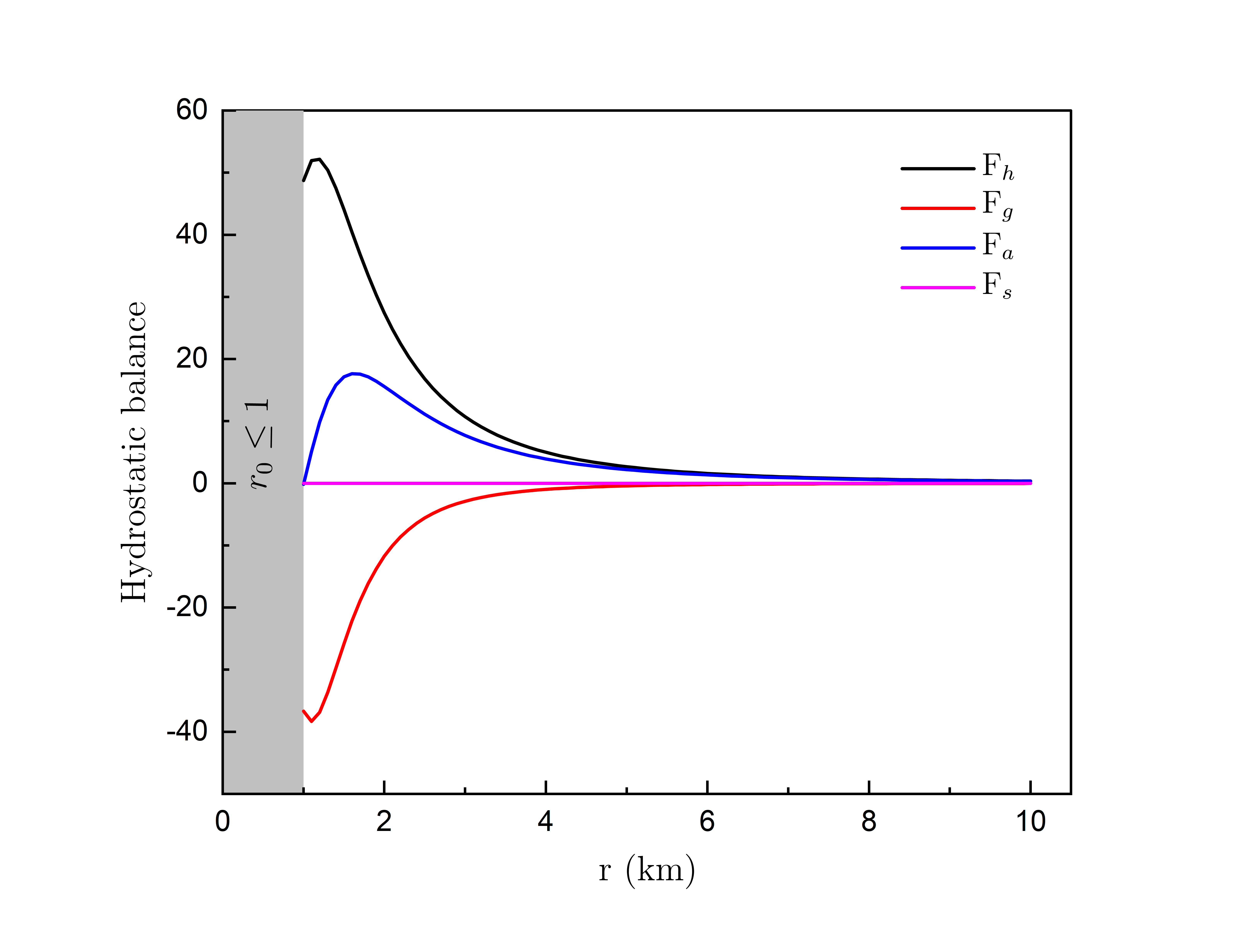}
    \caption{The hydrostatic balance of the structure under different forces.}
   \label{fig:hydro}
\end{figure}

From Fig. \ref{fig:hydro}, we can visualize that the hydrostatic condition is satisfied at a large distance thereby, concluding that the system is in
static equilibrium. The spin force is almost negligible for the adopted shape function.

\section{Conclusion}
With the help of Cartan’s equations of structure and Newmann-Penrose-Jogia-Griffiths formalism, the tetrad form of various curvature tensors like Riemannian tensor, Ricci tensor, etc. has been computed. We have found a tetrad form of energy-momentum tensor for Weysenhoff fluid. Using the conservation law of energy-momentum tensor the spin density has been derived in the form of red-shift function.\\
\begin{center}
\begin{table}[H]
    \centering
    \begin{tabular} { | c | c | c | c | }
    \hline
     Sr. No. & Quantities & Signs  \\
    \hline
     1& $\rho$ & $>0$ for $r>0.3$  \\
      \hline
   2 & $\rho+p_r$ & $>0$ for $r>0.3$  \\
   \hline
    3 & $\rho+p_t$ &$>0$ for $r>0.1$\\
     \hline
        4 & $\rho+p_r+2p_t$ & $>0$ for $r>0.1$ \\
     \hline
        5& $\rho-|p_r|$ & $>0$ for $r>0.3$ \\
     \hline
    6 & $\rho-|p_t|$ & $<0$ for all $r$ \\
     \hline
    \end{tabular}
    \caption{The overview of energy conditions}
    \label{tab:1}
\end{table}
\end{center}

The detailed analysis of EC for different ranges of throat radius are listed in Table \ref{tab:1}.
We studied a particular scenario in which the red-shift function is taken as $\frac{r_0}{r^n}$ and the shape function is $re^{1-\frac{r}{r_0}}$. In addition, geometric configurations and energy conditions are determined. From Table \ref{tab:1}, we can see that DEC is violated everywhere. With the use of exotic matter, which serves as the supporting matter for wormhole geometry in the framework of general relativity, traversable static wormholes may be possible and consequently, the WEC is also violated. However, the existence of such wormholes is possible in ECT in the absence of exotic matter. The energy conditions WEC and NEC are valid for $r>0.3$ and SEC is valid for $r>0.1$ as mentioned in the above table and Fig.3. For $r\leq0.3$ the energy conditions are violated and hence we get some exotic type matter in that region. Thus, in our case radius of the throat must be greater than 0.3 for an exotic-free wormhole. Our investigation into the stability of a wormhole within the framework of the Einstein-Cartan theory has provided valuable insights into the role of spin density in determining the equilibrium state of the wormhole structure. By utilizing the TOV equation, we were able to analyze the interplay of various forces like gravitational, hydrostatic, anisotropy, and spin forces acting on the wormhole. Stability analysis showed that the spin force has a negligible impact on the equilibrium of the wormhole for particular choice shape functions. This finding suggests that within the considered framework, the inclusion of spin density does not significantly affect the overall stability of the wormhole.

\section*{Acknowledgments}
SVS is thankful to the CSIR, India for providing financial support under CSIR Senior Research Fellowship (09/157(0059)/2021-EMR-I).

%\section*{References}

\bibliographystyle{ws-ijmpd}
\bibliography{sample}

\end{document}